\documentclass[twocolumn,prl,nofootinbib,superscriptaddress,showpacs]{revtex4}
\usepackage{graphicx}

\newcommand{\ie}{{\it i.e.}}

\newcommand{\fig}{Fig.}

\newcommand{\Ref}{Ref.}

\newcommand{\stheta}{\sin^22\theta_{13}}
\newcommand{\deltacp}{\delta_\mathrm{CP}}

\newcommand{\figu}[1]{\fig~\ref{fig:#1}}

\newcommand{\bi}{\begin{itemize}}
\newcommand{\ei}{\end{itemize}}

\begin{document}

\title{Upgraded experiments with super neutrino beams: Reach versus Exposure}
\author{V. Barger}
\author{Patrick Huber}
\affiliation{Department of Physics, University of Wisconsin, Madison, WI 53706, USA}
\author{Danny Marfatia}
\affiliation{Department of Physics and Astronomy, University of Kansas,
Lawrence, KS 66045, USA}
\author{Walter Winter}
\affiliation{Institut f{\"u}r theoretische Physik und Astrophysik,
Universit{\"a}t W{\"u}rzburg, D-97074 W{\"u}rzburg, Germany}

%\date{\today}

\begin{abstract}
\vspace*{0.1cm}
  We introduce {\it exposure} as a means to making balanced 
  comparisons of the sensitivities of long-baseline neutrino experiments 
  to a nonzero $\theta_{13}$, to CP violation and to the
  neutrino mass hierarchy. 
  We illustrate its use by comparing the sensitivities of possible 
  upgrades of superbeam experiments, namely NO$\nu$A*, T2KK and 
  experiments with wide band beams. 
  For the proposed exposures, we find the
  best nominal CP violation performance for T2KK. For equal exposures,
  a wide band beam experiment has  the best mass hierarchy performance. 
  The physics concept on which NO$\nu$A* is based
  has the best potential for discovering CP violation only for
  exposures above a threshold value.
\end{abstract}

%\keywords{}

\maketitle

{\bf{Introduction.}}  Extensive recent experimental exploration has
revealed that neutrinos are massive~\cite{rev}. This finding
necessitates the existence of physics beyond the Standard Model of
particle physics. Massive neutrinos may also have far-reaching
consequences for cosmology. They may shed light on the origin of the
baryon asymmetry in our universe and on why the universe is in an
accelerating phase in its expansion. It is therefore imperative that
the origin of neutrino masses be determined.

A plethora of neutrino mass models have been proposed and precise
knowledge of neutrino parameters is required to test them.
Specifically, the value of the mixing angle $\theta_{13}$ and the type
of mass hierarchy (\ie, whether $m_1, m_2 < m_3$, called the normal
hierarchy or $m_1, m_2 > m_3$, called the inverted hierarchy) will
help distinguish between models based on lepton flavor symmetries,
models with sequential right-handed neutrino dominance and more
ambitious models based on GUT symmetries~\cite{modelrev}.  A 
survey of 63 models that are consistent with current oscillation data
and have concrete predictions for $\theta_{13}$ found that half of
them predict $\sin^2 2\theta_{13}> 0.015$~\cite{albright}. GUT models
and models with right-handed neutrino dominance naturally yield a
normal hierarchy and a relatively large $\theta_{13}$ (although in a
few GUT models, an inverted hierarchy can be obtained with
fine-tuning). Models based on leptonic symmetries can easily
accommodate an inverted hierarchy and small $\theta_{13}$. Thus,
experimental establishment of an inverted hierarchy and small
$\theta_{13}$ would lend support to models based on leptonic
symmetries and reduce the interest in GUT models and models with
right-handed neutrino dominance. On the other hand, if $\theta_{13}$
is found to be large, distinguishing between the three different
classes of models will be difficult.  However, if in addition to a
large $\theta_{13}$, the hierarchy is found to be inverted, it will be
possible to exclude the subclass of SO(10) GUT models that employ
so-called lopsided mass matrices because they predict a normal
hierarchy.

Clearly, experiments with good sensitivity to $\theta_{13}$ and the
mass hierarchy are indispensable for sifting out a restricted class of
neutrino mass models. Precision measurements of deviations of the
atmospheric oscillation angle $\theta_{23}$ from $\pi/4$ are also
useful in distinguishing between models. The deviation from maximal
atmospheric mixing provides an excellent probe of how symmetry
breaking occurs in models based on leptonic symmetries.  The Dirac CP
phase $\deltacp$ in the neutrino mixing matrix may be related to the
CP violation required for leptogenesis~\cite{blanchet} 
(which is a direct consequence
of the seesaw mechanism) and it may therefore be possible to test both
the seesaw and the origin of the baryon asymmetry in our universe by
measuring this CP phase.

If neutrinos do not have approximately degenerate masses, the
sensitivity of experiments seeking to detect neutrinoless double beta
decay (thereby confirming that neutrinos are Majorana particles), is
strongly impacted by whether the mass hierarchy is normal or inverted.

Long-baseline neutrino experiments offer the only way to establish a
nonzero $\theta_{13}$, to determine the mass hierarchy and to detect
neutrino CP violation. There are two strategies being considered for a
future experimental program, with combinations of different types of
neutrino beams and detector technologies. Off-axis beams have a narrow
beam energy, permitting a counting experiment at an oscillation
maximum with low background. Wide band beams have a higher flux and
allow an experiment that utilizes spectral energy information, but
requires sophisticated detectors with good energy
resolution and neutral-current rejection to reduce backgrounds.

The Tokai-to-Kamioka (T2K) experiment~\cite{jhfsk} will use an
off-axis beam. The proposed NuMI Off-axis $\nu_e$ Appearance
(NO$\nu$A) experiment~\cite{nova} (and its second phase) and the
Tokai-to-Kamioka-and-Korea (T2KK) extension~\cite{t2kk} of the T2K
experiment also plan to employ off-axis beams.  Recently, a wide band beam (WBB)
experiment has been advocated~\cite{wide-band}, the virtues of which have been
investigated in Ref.~\cite{wbb}. With the looming possibility of a
Deep Underground Science and Engineering Laboratory
(DUSEL)~\cite{dusel} in the U.S., and its capacity to house very large
detectors, it is timely to evaluate the relative merits of the two
experimental approaches with upgraded superbeams.

So far, the experimental options and assumptions made in analyses have
been so diverse that an objective comparison is not possible. For
example, one experiment may seem to have greater sensitivity simply
because the exposure assumed is much larger than that of another.

We carry out a technically comprehensive study with a realistic
treatment of systematic errors, correlations and
degeneracies~\cite{Barger:2001yr}.  Our goal is to clarify the physics
reach of the different proposals by analyzing them on an equal
footing. We present the sensitivities of the experiments to a nonzero
$\theta_{13}$, the mass hierarchy and to CP violation {\it as a
  function of exposure} so that merits of the different experimental
techniques are evident.

\begin{table*}
\begin{tabular}{lccccccccr}
\hline
Setup & POT $\nu$/yr & $t_{\nu}$ [yr] & POT $\bar{\nu}$/yr & $t_{\bar{\nu}}$
[yr] & $P_\mathrm{Target}$ [MW] & $L$ [km] & Detector technology &
$m_{\mathrm{Det}}$ [kt] & $\mathcal{L}$ [$\mathrm{Mt \, MW \, 10^7 \, s}$]\\
\hline
%NO$\nu$A-I & $10 \cdot 10^{20}$ & 3+3 & $10 \cdot 10^{20}$ & 3+3 & 1 & 810 &
%TASD & 25 & 0.51 \\
NO$\nu$A* & $10 \cdot 10^{20}$ & 3 & $10 \cdot 10^{20}$ & 3 & 1.13 & 810 &
LArTPC & 100 & 1.15 \\
%NO$\nu$A (I+II) & & & & & & & & & 1.53 \\
WBB+WC & $22.5 \cdot 10^{20}$ & 5 & $45 \cdot 10^{20}$ & 5 & 1 ($\nu$), 2
($\bar{\nu})$& 1290 & Water Cherenkov & 300 & 7.65 \\
WBB+LAr & $22.5 \cdot 10^{20}$ & 5 & $45 \cdot 10^{20}$ & 5 & 1 ($\nu$), 2
($\bar{\nu}$) & 1290 & LArTPC & 100 & 2.55 \\
T2KK & $52 \cdot 10^{20}$ & 4 & $52 \cdot 10^{20}$ & 4 & 4 &
295+1050 & Water Cherenkov & 270+270 & 17.28
\\
\hline
\end{tabular}
\caption{\label{tab:setups} Setups considered, numbers of protons on target
per year (POT/yr) for the neutrino and antineutrino running modes,
running times in which these be achieved, corresponding target power
$P_\mathrm{Target}$, baselines $L$, detector technology, detector mass
$m_{\mathrm{Det}}$, and exposure $\mathcal{L}$.}
\end{table*}

{\bf{Experimental setups and analysis techniques.}}  We use the GLoBES
software~\cite{globes} for our simulations.  Table~\ref{tab:setups}
displays parameters of the experiments. 

Our simulation of NO$\nu$A phase~II, which we call NO$\nu$A*, is
based upon the proposal~\cite{nova} and recent studies on the
performance of a Liquid Argon Time Projection Chamber 
(LArTPC)~\cite{Flemming:2006}. We
assume NO$\nu$A* (3 years $\nu$ and 3 years $\bar\nu$) with a
100~kt LArTPC, which has a 0.8 signal efficiency and only beam
intrinsic $\nu_e$ and $\bar\nu_e$ backgrounds. We split the event
sample into quasi-elastic (QE) events with $5\%$ energy resolution and
the non-QE charged current events with $20\%$ energy resolution. We
have carried out a dedicated optimization study in baseline versus
off-axis angle plane whose details can be found in Ref.~\cite{Barger:prep}.
We find that the best location for all measurements is the
Ash River site (12~km off-axis at $L=810$~km) where NO$\nu$A phase~I
is located. None of the alternative sites such as in \Ref~\cite{olga}
performs as well as Ash River.  This result holds even if NO$\nu$A
phase~I data is taken into account.

For the WBB experiments, we use the simulation from \Ref~\cite{wbb}
which uses neutrino spectra obtained from 28~GeV protons and a 200 m
long decay tunnel, 
and choose the Fermilab-Homestake baseline $L=1290 \, \mathrm{km}$ for
reference.  We consider two possible detector technologies: A $300 \,
\mathrm{kt}$ water Cherenkov detector and a $100 \, \mathrm{kt}$
liquid argon TPC. We assume that five years of neutrino running with a
1~MW beam will be followed by five years of running with a 2~MW beam.

For the NO$\nu$A* and WBB setups,
 we use a systematic uncertainty of $5\%$ on both signal
and background, uncorrelated between neutrino and antineutrino channels.

For our T2KK simulation, we employ the values from \Ref~\cite{t2kk}
with a 2.5$^\circ$ off-axis beam. Our simulation is based upon the
analysis of the Tokai-to-HyperKamiokande experiment in
\Ref~\cite{Huber:2002mx}, \ie, we use the spectral information for
quasi-elastic (QE) events, and the total event rate for all charged
current (CC) events. 
We include 5\% signal and background errors, as
well as a 5\% background energy calibration error which are correlated
between the two detectors in Japan and Korea, but uncorrelated between
the neutrino and antineutrino channels. 

We adopt $\Delta m_{21}^2 = +8 \cdot
10^{-5} \, \mathrm{eV}^2$, $\Delta m_{31}^2 = +2.5 \cdot 10^{-3} \,
\mathrm{eV}^2$, $\sin^2 \theta_{12} =0.3$, $\sin^2 \theta_{23} =0.5$ for the
oscillation parameters. We assume that the atmospheric oscillation parameters are
measured to 10\%, the solar parameters are
measured to 4\%, and the matter density along the baseline is known to
5\%. We include all correlations and degeneracies in the analysis.
Details of our simulations are presented in \Ref~\cite{Barger:prep}. Since
we present sensitivities for each of the three 
performance indicators separately, we use $\chi^2$ distributions for
one degree of freedom. 

{\bf{Results.}}
\begin{figure*}
\begin{center}
\includegraphics[width=\textwidth]{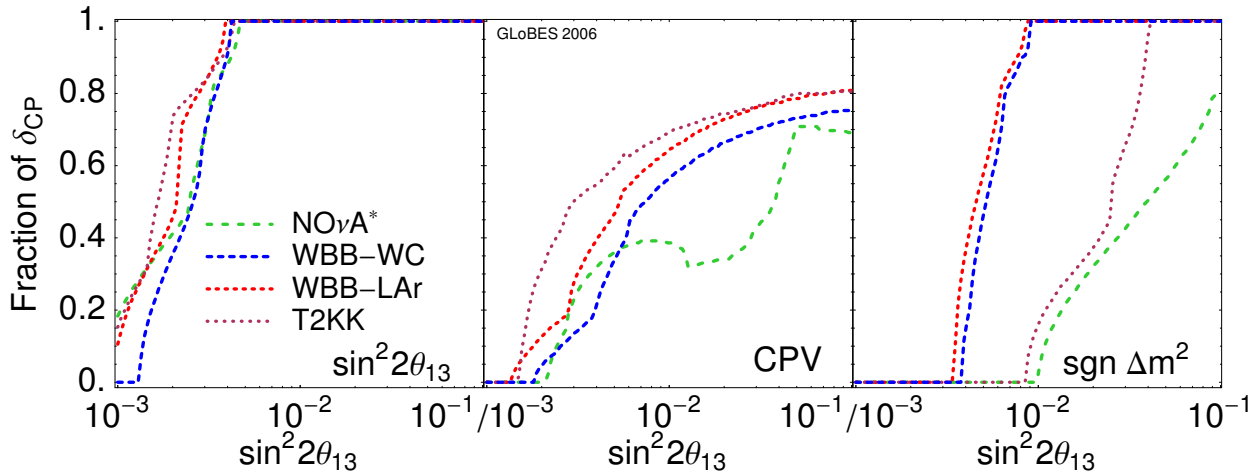}
\end{center}
\caption{\label{fig:comp} Comparison of superbeam upgrades in the
configurations
of Table~\ref{tab:setups} at the $3 \sigma$ C.L. The plots show the
discovery reaches
for a nonzero $\stheta$, CP violation,
and the normal hierarchy.
The ``fraction of $\deltacp$'', quantifies
the fraction of all
(true) values of $\deltacp$
for which the corresponding quantity can be measured.}
\end{figure*}
In~\figu{comp} we show the comparison of superbeam upgrades in the
configurations of Table~\ref{tab:setups} for the $\stheta$, CP
violation, and normal hierarchy discovery reaches.  This comparison
illustrates the absolute physics potentials of the planned experiments. 
Interestingly, the optimal physics performance
depends on the performance indicator. The $\stheta \neq 0$ discovery
reaches are very similar for all the experiments.  T2KK has the best
CP violation potential.  The WBB experiments can detect the mass
hierarchy down to $\stheta \simeq 10^{-2}$ for all values of
$\deltacp$, which makes them the best upgrade for the mass hierarchy
(as a result of their long baseline and high energy and consequently
strong matter effects~\cite{bpww}). However, this figure does not permit a
balanced assessment of which experiment is the best physics concept
because of the very different assumptions for the luminosities in each
proposed experiment.

In order to make an unbiased comparison of the physics potentials of
the experimental setups we consider their sensitivities as functions
of {\it {exposure}} which we define to be $\mathcal{L}=$ detector mass
[Mt] $\times$ target power [MW] $\times$ running time [$10^7$~s]. The
target power represents the bottleneck in technological difficulty.
Note that instead of the running time in years, the exposure uses the
actual available time of the accelerator for the neutrino experiment.
For NO$\nu$A* and the WBB, we use $1.7 \cdot 10^7$ seconds uptime per
year, and for T2KK, we use $10^7$ seconds uptime per year (as
anticipated in the corresponding documents). Note that this definition
does not account for the level of sophistication of different detector
technologies, but it will allow for an identification of the
break-even point of the detector cost. We show the exposure for the
discussed experiments in the last column of Table~\ref{tab:setups}.
It is evident that NO$\nu$A* has the lowest exposure, whereas T2KK has
the highest. While we will show a normalized comparison of the
experiments based on the exposure, there may be
other issues, such as robustness of systematics and a different
experiment optimization that may modify the conclusions. We will
discuss these issues elsewhere~\cite{Barger:prep}.

%\section{Results II}
% Results

\begin{figure}
\begin{center}
\includegraphics[width=\columnwidth]{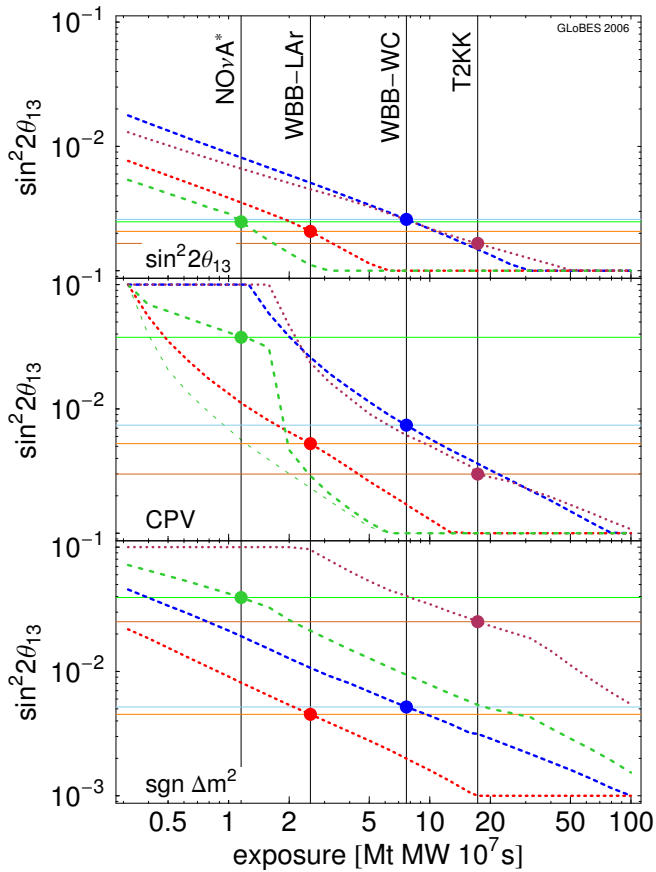} 
%\\[0.2cm]
%\includegraphics[width=\columnwidth]{cpv-lumi} \\[0.2cm]
%\includegraphics[width=\columnwidth]{sgn-lumi}
\end{center}
\caption{\label{fig:lumi} The discovery reaches (at the $3 \sigma$ C.L.)
  for nonzero $\stheta$, CP violation, and the normal hierarchy as
  functions of exposure.  The line types are the same as in
  \figu{comp} except that the light curve in the CPV panel corresponds to 
  the sensitivity of NO$\nu$A* under the assumption that the mass hierarchy 
  is known to be normal. The vertical lines mark the proposed luminosities as
  listed in Table~\ref{tab:setups}.  The curves correspond to a
  fraction of $\deltacp$ of 0.5, \ie, the median of the distribution.
  This means that the performance will be better for 50\% of all cases
  of $\deltacp$ and worse for 50\% of all cases of $\deltacp$; it is
  sometimes referred to as the ``typical value of $\deltacp$''.}
\end{figure}

In \figu{lumi} we show the discovery reaches for $\stheta$, CP
violation, and normal mass hierarchy versus the exposure for a
fraction of $\deltacp$ of 0.5 (see figure caption).  
The NO$\nu$A*
curves for $\stheta$ and CP violation discoveries are
lower than the ones of the other experiments for exposures above 
2~Mt~MW~10$^7$s, 
whereas the curves for the WBB experiments are lower (for any exposure) 
than any other
curve for the mass hierarchy discovery. If all experiments were
operated at the same exposure, these experiments would yield the best
results.
%However, for CP violation, the high exposure makes
%T2KK the best absolute experiment (as marked by the dot on the T2KK curve.
All the curves scale relatively smoothly as a function of exposure
except the CP violation curve for NO$\nu$A*. The bump-like feature
is solely due to the interplay of CP effects and the mass hierarchy 
and is called $\pi$-transit~\cite{Huber:2002mx}.{\footnote{$\pi$-transit
degrades the sensitivity to CP violation by a parametric conspiracy
which allows to fit data which was generated for a CP violating value
of $\delta$, with $\delta=\pi$ and the wrong hierarchy. 
The occurrence of this effect is tied to a certain, experiment specific
range of $\stheta$. When the median of the CP fraction moves into that
range of $\stheta$ a bump is observed. The bump occurs for any intermediate
value of the CP fraction chosen for a plot like \figu{lumi}.}} 
A further luminosity
increase could enhance the NO$\nu$A* potential for CP violation
considerably by enabling the resolution of degeneracies at this
confidence level; see the light curve in the CPV panel which is made
under the assumption that the hierarchy is known to be normal. 
 The other setups are relatively insensitive to
small variations in exposure. For CP violation, the WBB-WC and T2KK
concepts are more or less equivalent since the curves almost overlap.
The WBB-WC and the T2KK curves intersect at some points. These
intersections limit the exposure ranges in which one experiment
dominates the other.  For example, for $\stheta$, T2KK plans to
operate with an exposure for which the WBB-WC concept would perform
slightly better, whereas a significantly lower exposure would make
T2KK the more sensitive experiment.  Finally, one can read off the
break-even point between the water Cherenkov and liquid
argon-technologies in WBB experiments.  For example, for $\stheta$,
the water Cherenkov and liquid argon technologies are separated by
about a factor of 4 in exposure, which means that liquid argon is
the choice of technology if the cost per kt of liquid argon is smaller
than the cost for 4~kt water.  Note that the corresponding
sensitivities to CP violation and the mass hierarchy are quite
similar.

{\bf{Summary and conclusions.}} It is crucial that the mixing angle
$\theta_{13}$, the nature of the neutrino mass hierarchy and whether
CP is violated in the neutrino sector, be determined to complete the
parameter set that defines the neutrino mass matrix. This program is
of fundamental value for understanding the origin of neutrino masses
and for selecting between neutrino mass models.

In the not-too-distant future, the planning stage for long-baseline
neutrino experiments with super neutrino beams and large detectors
will end. We have provided the first analysis of various experimental
configurations on an equal-footing by expressing their sensitivities
as functions of exposure. 
By enabling a balanced comparison, our study
identifies which physics concept is optimal for which measurement.  If
a large liquid argon TPC can become a reality, our analysis indicates
that with an adequate increase in exposure, an upgraded NO$\nu$A
experiment like NO$\nu$A* has better sensitivity to a 
nonzero $\theta_{13}$ and
to CP violation than previous estimates suggested. The longer
baselines planned for experiments with wide band beams offer better
sensitivity to the mass hierarchy. 

The power of assessing sensitivities as functions of exposure 
is manifest in the CPV sensitivity of NO$\nu$A*. 
%it became evident that if the nominal
%sensitivity of NO$\nu$A* is doubled, the sensitivity to CPV improves
%by an order of magnitude. 
This method is applicable to all long-baseline
neutrino experiments where it may provide crucial insights into
optimal experimental configurations. Since exposure is
a measure of the integrated luminosity, it can also be used in comparative
evaluations of other kinds of experiments.

%\vspace*{0.05cm}
{\it Acknowledgments.} 
We thank M. Bishai, M. Dierckxsens and M. Diwan 
for useful discussions.
This research was supported by the U.S.
DoE under Grants No. DE-FG02-95ER40896 and
DE-FG02-04ER41308, by the NSF under CAREER Award No. PHY-0544278, by
the State of Kansas through KTEC, by the KU GRF Program, and by the 
Emmy Noether Program of the Deutsche Forschungsgemeinschaft.
%Computations were performed on facilities supported by the NSF under
%Grants No. EIA-032078 (GLOW), PHY-0516857 (CMS Research Program
%subcontract from UCLA), and PHY-0533280 (DISUN), and by the WARF.

\end{document}